\documentclass{ifacconf}
\pdfoutput=1
\usepackage{graphicx}      
\usepackage{natbib}        
\usepackage{amsmath} 
\usepackage{amssymb}  

\usepackage{eso-pic}

\begin{document}

\AddToShipoutPictureBG*{%
\put(0,20){
\hspace*{\dimexpr0.075\paperwidth\relax}
\parbox{.84\paperwidth}{\footnotesize \copyright2023 the authors. This work has been accepted to IFAC for publication  under a Creative Commons Licence CC-BY-NC-ND}%
}}
\begin{frontmatter}

\title{Fast or Cheap: Time and Energy Optimal Control of Ship-to-Shore Cranes} 

\thanks[footnoteinfo]{This work was supported by VINNOVA Competence Center LINK-SIC.}

\author[First]{Filipe Marques Barbosa} 
\author[First]{Anton Kullberg} 
\author[First]{Johan Löfberg}

\address[First]{Division of Automatic Control, Linköping University, Sweden \\e-mail: \{filipe.barbosa, anton.kullberg, johan.lofberg\}@liu.se.}

\begin{abstract}                
This paper addresses the trade-off between time- and energy-efficiency for the problem of loading and unloading a ship. Container height constraints and energy consumption and regeneration are dealt with. We build upon a previous work that introduced a coordinate system suitable to deal with container avoidance constraints and incorporate the energy related modeling. In addition to changing the coordinate system, standard epigraph reformulations result in an optimal control problem with improved numerical properties. The trade-off is dealt with through the use of weighting of the total time and energy consumption in the cost function. An illustrative example is provided, demonstrating that the energy consumption can be substantially reduced while retaining approximately the same loading time.
\end{abstract}


\end{frontmatter}

\section{Introduction}

\begin{figure}[bt]
	\centering
	\includegraphics[width=.88\columnwidth]{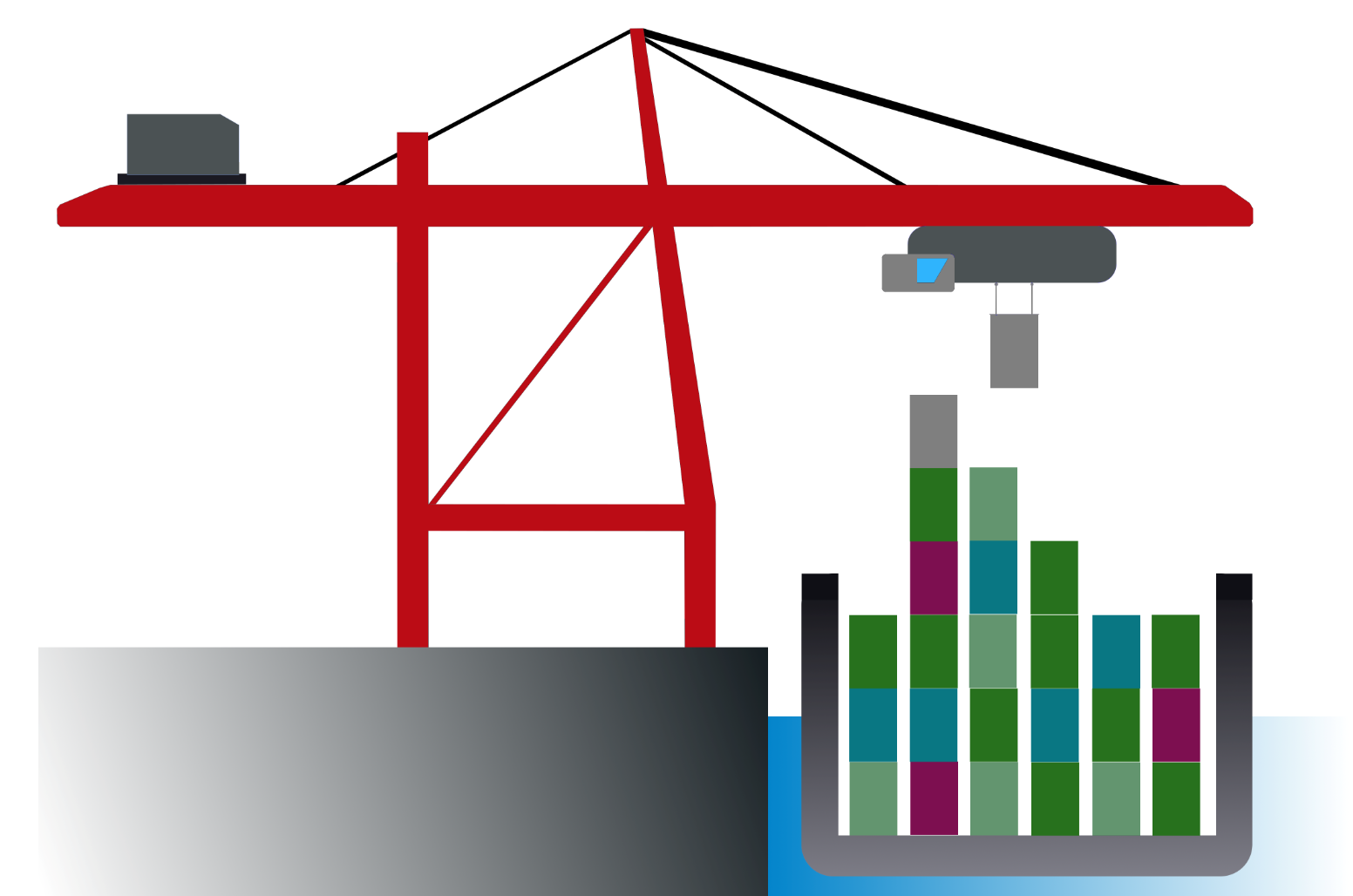}
	\caption{A ship-to-shore crane schematic.}
	\label{fig:sts}
\end{figure}

Shipping containers remarkably improved the productivity of ports and revolutionized the global commerce. Moreover, the speed at which a ship is unloaded and loaded again, as well as the energy cost per container, are important performance indicators of a port's operations \citep{wilmsmeier2016energy,Kreuzer2014,Sakawa1982}. Thus, the efficiency of port operations depends deeply on the performance of ship-to-shore (STS) cranes, see Fig.~\ref{fig:sts}. They are responsible for the critical task of safely unloading and loading the ship \citep{Kreuzer2014,Sakawa1982,Arena2015}. Therefore, making their operations faster is crucial for the performance of a port.

Optimizing loading time can be done in various ways. Time can be minimized implicitly by reducing the sway of the container, which has been considered in, e.g., \cite{Kreuzer2014,kim2004anti}. To explicitly minimize the loading time, time-optimal control is a natural approach. In such approaches, the objective is to minimize the total time for the payload to go from an initial to a final position, see e.g., \cite{Sakawa1982,Auernig1987,Chen2016}. Additionally, a combination of both sway and time minimization can be used, e.g., \cite{AlGarni1995,DaCruz2012}.

However, simply optimizing loading and unloading time may cause the actuators to operate close to saturation. This in turn leads to excessive energy consumption and severe stress on the actuators themselves, which can increase maintenance requirements and costs. This said, a natural way of addressing this problem is to minimize the energy consumed to move the payload along the loading site.

Energy consumption of overhead cranes have been analyzed, and efficient solutions have been presented in, e.g., \cite{Kosucki2020,Kosucki2017,Wu2014}. However, only minimizing energy leads to increased loading time, reducing the productivity of the port. Thus, a trade-off between time optimality and energy efficiency is necessary, see e.g., \cite{wang2021energy}.  Moreover, STS cranes may have energy regeneration systems, allowing them to recover energy during the loading process \citep{kusakana2021optimal,Zhao2016}. Thus, this feature can be taken into account when formulating time- and energy-optimal solutions.

However, most of the approaches in the literature still do not consider the different height of container stacks and their setup along the loading site. Though some authors do include hoisting, container stack configuration is nonetheless not considered, see for instance \cite{DaCruz2012,Kosucki2020}. Moreover, when dealing with energy consumption in STS cranes, the hoisting mechanism is the primary energy consumer and thus should certainly be taken into account.

In this paper, we propose an optimal control approach to address the trade-off between time- and energy-efficiency for the problem of loading and unloading a ship subject to container avoidance constraints. We build upon an earlier time-optimal approach that addresses the container avoidance constraints, presented by \cite{barbosa2022time}. This was done by reparametrizing the problem in spatial coordinates, which turns the avoidance problem into simple linear bound constraints. Here, we extend this framework to include energy consumption and regeneration of the trolley and hoisting mechanisms. The problem is formulated to avoid the non-smoothness in the cost function, caused by the inclusion of energy terms. This in turn results in an optimal control problem with improved numerical properties. The trade-off is dealt with in the cost function through a weighting of the total time and energy consumption. Finally, we show that the energy consumption can be substantially reduced while retaining approximately the same loading time.

\section{Model and Problem Formulation}\label{sec:modelling_and_original}

To trade-off between time and energy consumption in the optimal control of an STS crane subject to container avoidance constraints, we first derive a state-space model. Then we address the energy consumption accounting for energy regeneration and finally an optimization problem is formulated. 



\subsection{Modeling}\label{sec:modelling}

\begin{figure}[t]
	\centering
	\includegraphics[width=.9\columnwidth]{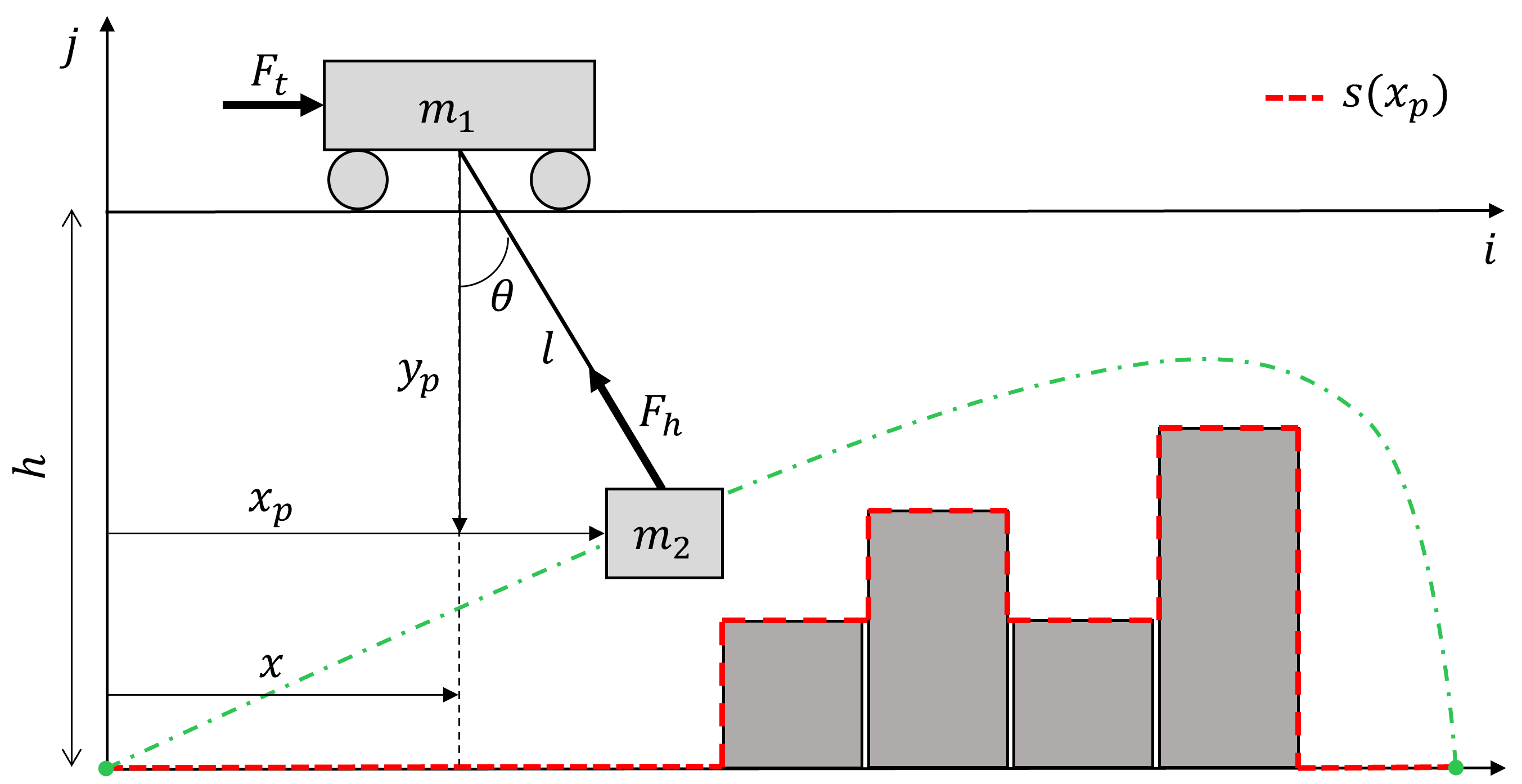}
	\caption{Two-dimensional schematic representing the trolley and payload motions, with the container stacks acting as constraints for the payload trajectory.}
	\label{fig:schematic}
\end{figure}

Since STS cranes load one row of containers at a time, the trolley movement is restricted to one dimension. Therefore, a 2D-model of the dynamics suffices. Thus, a cart-pendulum with a single-rope hoisting mechanism can be used to represent the crane's dynamics. Here the dynamics of the actuators themselves are not considered, and the payload is treated as a point mass. This way, the generalized coordinates are the length of the hoisting rope $l(t)$, the sway $\theta(t)$, and the payload coordinates $x_p(t)$ and $y_p(t)$. The forces applied to the trolley $F_T(t)$ and to the hoisting mechanism $F_H(t)$ are the control inputs, where subscript $T$ and $H$ refer to the trolley and hoisting mechanism, respectively. See Fig.~\ref{fig:schematic} for a schematic. We refer to \cite{barbosa2022time} for the equations of motion and restate the nonlinear state equations here. The state variables are $x=[x_p,\dot{x}_p,y_p,\dot{y}_p,l,\dot{l},\theta,\dot{\theta}]^T$ and the model is given by
\begin{align}
		\dot{x}_{1} &= x_{2}\nonumber\\
		\dot{x}_{2} &= -(u_{2}\sin(x_{7}))/m_{2} \nonumber\\
		\dot{x}_{3} &= x_{4}\nonumber\\
		\dot{x}_{4} &= -(u_{2}\cos(x_{7}))/m_{2}+g \label{eq:original-state-eq}\\
		\dot{x}_{5} &= x_{6}\nonumber\\
		\dot{x}_{6} &= x_{5}x_{8}^{2}\!+\!g\!\cos(x_{7})\!-\!u_{2}/m_{2}\!-\!\sin(x_{7})(u_{1}\!+\!u_{2}\sin(x_{7}))/m_{1}\nonumber\\
		\dot{x}_{7} &= x_{8}\nonumber\\
		\dot{x}_{8} &= -(2x_{6}x_{8}\!+\!g\!\sin(x_{7})\!+\!\cos(x_{7})\!(u_{1}\!+\!u_{2}\sin(x_{7}))/m_{1})/x_{5},\nonumber
\end{align}
with $m_1$ and $m_2$ respectively the trolley and payload masses, $g$ the acceleration of gravity, and the control inputs $u_{1} = F_{T}$ and $u_{2}=F_{H}$.


\subsection{Energy consumption and regeneration}

The energy consumed to move the trolley or to  hoist the payload can be expressed as
\begin{equation}
\centering
    E(t) = \int_{0}^{t} F(\tau)v(\tau) d\tau = \int_{0}^{t} P(\tau) d\tau,
\end{equation}
where $F$ is the driving force, $v$ is the velocity of the object on which the force is applied, and consequently $P$ is the power applied to the respective subsystem. 

In this paper, we want to study the case when we can recover part of the energy that has been fed into the system by using regenerative braking on the trolley and lowering of the payload. A simple model of this regeneration will be used to study the conceptual properties of solutions. Thus, when a system consumes energy, $P$ is positive, meaning that the source is providing energy to it. When energy is recovered, $P$ is negative and energy is being pulled back to the source with efficiency $0\leq \gamma\leq 1$. Hence, the power flow of a subsystem with energy recovery is given by $\max(P,\gamma P)$ and its energy consumption by
\begin{equation}
\label{eq:opt-energy}
E(t) = \int_{0}^{t}\max(P(\tau), \gamma P(\tau))d\tau.
\end{equation}

In STS crane operations, the energy optimal formulation is given by the minimization of the total energy spent to move the trolley $E_T$ and to hoist the payload along the loading site $E_H$, i.e., \eqref{eq:opt-energy} is written as
\begin{multline}
	\label{eq:opt-energy-hoist-trolley}
	E(t)=\underbrace{\int_{0}^{t_f}\max(\!P_T(t),\gamma_TP_T(t))dt}_{E_T}\\
	+\underbrace{\int_{0}^{t_f}\max(P_H(t),\gamma_HP_H(t))dt}_{E_H}
\end{multline}

with $0 \le \gamma_T,\gamma_H \le 1$ the efficiency of energy recovery of the trolley and hoisting mechanism, respectively. Furthermore, 
\begin{equation}
\label{eq:power-time}
    \begin{aligned}
    P_T(t) &= F_T(t)\dot{x}_T(t)\\
    &= F_T(t)(\dot{x}_p(t)\!-\!\sin(\theta(t))\dot{l}(t)\!-\!l(t)\cos(\theta(t))\dot{\theta}(t)) \\
    P_H(t) &= -F_H(t)\dot{l}(t),
    \end{aligned}
\end{equation}
where $P_T$ and $P_H$ correspond to the power used in the trolley and in the hoisting mechanism, respectively, and $\dot{x}_T(t)$ is obtained from direct differentiation of the trolley position $x_T(t) = x_p(t)\!-\!\sin(\theta(t))l(t)$
in Fig.~\ref{fig:schematic}. 

\subsection{Time- and energy-optimal control}\label{sec:orignal_prob}
To account for energy consumption, terms of \eqref{eq:opt-energy} would be included in the cost function. However, we argue that a better formulation is to avoid a non-smooth integrand in the cost function, which will lead to non-smooth terms in a discretized objective. We therefore introduce an auxiliary variable $z(t)$ acting as an upper bound in an epigraph formulation, thus leading conceptually to
\begin{equation}
\label{eq:opt-energy_ref}
\begin{aligned}
& \underset{u}{\text{min }}
&& \int_{0}^{t_f}z(t)dt\\
& \text{subject to}
&& z(t)\ge P(t)\\
&&& z(t)\ge \gamma P(t).\\
\end{aligned}
\end{equation}
Note that at optimality $z(t)\ge P(t)$ is tight when the system is consuming energy, and $z(t)\ge \gamma P(t)$ is tight when energy is being recovered.
Thus, the formulation in \eqref{eq:opt-energy_ref} can be applied to \eqref{eq:opt-energy-hoist-trolley} with the corresponding auxiliary variables, i.e., $z_T(t)$ and $z_H(t)$ acting as upper bounds in the epigraph formulation. Allowing for a compromise between time- and energy-optimality through a parameter $0\leq  \alpha \leq 1$, the optimal control problem is given by
\begin{align}
\label{eq:original-formulation}
&\begin{aligned}
& \min_u
& & J = \alpha\int_{0}^{t_f}dt\!+\!(1\!-\!\alpha)\!\int_{0}^{t_f}\!z_T(t)\!+\!z_H(t)\, dt \\
& \text{subject to} && \\
\end{aligned}\nonumber\\
&\begin{aligned}
&\; \dot{x}(t) = f(t,x(t),u(t)), &&\\
&\; x(0) = x_{0}, && z_T(t) \ge P_T(t),\\
&\; x(t_f) = x_{t_f}, && z_T(t) \ge \gamma_TP_T(t),\\
&\; g(x(t)) \le 0, && z_H(t) \ge P_H(t),\\ 
&\; u_{min}(t) \le u(t) \le u_{max}(t), && z_H(t) \ge \gamma_HP_H(t).
\end{aligned}
\end{align}

Thus, \eqref{eq:original-formulation} allows for a trade-off between time- and energy-optimality. However, the problem of avoiding collision with container stacks still remains, which will be addressed now.
\section{Problem reformulation}

When transferring the payload, an STS crane must avoid crashing it into stacks of already loaded containers. This imposes constraints on the payload height $y_p(x_p)$. Hence, a function $s(x_p)$ is conceptually introduced (see Fig.~\ref{fig:schematic}) to represent the profile of already loaded container stacks along the loading site, yielding the constraints on the payload height as 
\begin{equation}
0 \le y_{p}(x_{p}) \le h-s(x_{p}).
\label{eq:height_constraint_general}
\end{equation}
However, representing $s(x_p)$ is challenging when the independent variable in the optimal control problem is time $t$. When solving the optimization problem using a numerical solver, discretization using $s(x_p(t^k))$ in discretization points $t^k$ are required, and $s(x_p)$ will generally be discontinuous, nonlinear and non-convex \cite{barbosa2022time}. To resolve these problems, we perform a variable change and utilize a reformulation of the optimization problem, see \cite{barbosa2022time} for more details.

\subsection{Reparametrization from time to space}\label{sec:reformulation}

Time discretization will force the use of an explicit functional representation $s(x_p)$ of the geometric constraints and lead to issues, as mentioned above. On the other hand, these constraints turn out to be trivial to represent if the discretization is performed along the spatial dimension $x_p$. Therefore, we change the integration variable and define the dynamics in terms of the spatial position $x_{p}$ instead of time $t$, similarly to \cite{Verscheure2009}.

Now consider the first dynamic equation in \eqref{eq:original-state-eq}, which describes the payload velocity along $x_p$, then
\begin{equation}
\label{eq:change_state}
\dot{x}_1=\frac{dx_1}{dt} = x_2 \implies dt = \frac{dx_1}{x_2} \implies \frac{dt}{dx_1} = \frac{1}{x_2}.
\end{equation}
Thus, the cost function in \eqref{eq:original-formulation} is rewritten as 
\begin{equation}
	\label{eq:obj-reformulation}
	J\!=\alpha\!\int_{x_{p_0}}^{x_{p_f}}\!\frac{1}{x_2}dx_p\!+(1\!-\!\alpha)\!\int_{x_{p_0}}^{x_{p_f}}\!\frac{z_T(x_p)\!+\!z_H(x_p)}{x_2}\,dx_p.
\end{equation}
With this change of variables and forthcoming additional reformulations, the dynamics is expressed in the payload position $x_p$ and \eqref{eq:obj-reformulation} is the cost function to be minimized.

Following the transformation in \eqref{eq:change_state}, a new state vector  $x=[t,\dot{x}_p,y_p,\dot{y}_p,l,\dot{l},\theta, \dot{\theta}]^T$ is defined for the problem.

\textit{Remark:} To avoid confusion, we use $x^\prime=dx/dx_p$ for derivatives with respect to $x_p$, and $\dot{x} = dx/dt$ for derivatives with respect to time $t$.

Furthermore, since time $t$ is now a state variable, and the variable in which the dynamics is expressed is $x_p$, the following identification is made
\begin{equation}
	\label{eq:identifications}
	\setlength{\arraycolsep}{1pt}
	x_{1}\leftarrow{t},~x^\prime_{j}\leftarrow{\frac{dx_{j}}{dx_{p}}},~ j = 1, \ldots, n,
\end{equation}
where $j$ indices the state variables in the new state vector and $n$ is the system's dimension. Therefore, the state equations in (\ref{eq:original-state-eq}) become
\begin{align}
x_{2}x^\prime_{1}\!&=\!1\nonumber\\
x_{2}x^\prime_{2}\!&=\!-(u_{2}\sin(x_{7}))/m_{2}\nonumber\\
x_{2}x^\prime_{3}\!&=\!x_{4}\nonumber\\
x_{2}x^\prime_{4}\!&=\!-(u_{2}\cos(x_{7}))/m_{2}+g\label{eq:reformulated-state-eq}\\
x_{2}x^\prime_{5}\!&=\!x_{6}\nonumber\\
x_{2}x^\prime_{6}\!&=\!x_{5}x_{8}^{2}\!+\!g\!\cos(x_{7})\!-\!u_{2}\!/\!m_{2}\!-\!\sin(x_{7})\!(u_{1}\!+\!u_{2}\!\sin(x_{7}))\!/\!m_{1}\nonumber\\ 
x_{2}x^\prime_{7}\!&=\!x_{8}\nonumber\\
x_{2}x^\prime_{8}\!&=\!-(2x_{6}x_{8}\!+\!g\!\sin(x_{7})\!+\!\cos(x_{7})\!(u_{1}\!+\!u_{2}\!\sin(x_{7})\!)\!/\!m_{1})\!/\!x_{5}.\nonumber 
\end{align}

This means that with representation and discretization in $x_p$, the stack constraints can be represented by specific function values of $s(x_p)$ at the discretization points. Then, the constraints on the payload height, in a numerical discretization, is simply given by
\begin{equation}
	0 \le y_{p}(x_{p}^{k}) \le h-s(x_{p}^{k}).
	\label{eq:height_constraint_space}
\end{equation}
Now $s(x_p^k)$ is the stack height at a specific discretization point $x_p^k$. Therefore, $s(x_p^k)$ is a constant, which leads to simple bound constraints on $y_p$, in contrast to \eqref{eq:height_constraint_general}.

\subsection{Problem formulation in spatial coordinate}


With the transformation \eqref{eq:change_state}, \eqref{eq:obj-reformulation} is a natural choice of the cost function. However, this unnecessarily adds the risk of numerical problems in a solver as it involves a division by a variable that is typically zero at the end-points. To avoid these possibly problematic Lagrange terms in the optimal control problem, we first note that the total time simply evaluates to $t_f$, which is the final value of the first state as $x_{1}(x_{p})=t(x_p)$. Additionally, we introduce two new states to \eqref{eq:reformulated-state-eq}
\begin{subequations}
\label{eq:mayer_states}
\begin{align}
	    &x_{2}x^\prime_{9} = z_T(x_p)  \text{ and }\\
	    &x_{2}x^\prime_{10} = z_H(x_p). 
\end{align}
\end{subequations}
Thus, the total energy consumed to move the trolley and to hoisting the payload is given by evaluating \eqref{eq:mayer_states} to $x_{p_{f}}$ as
\begin{subequations}
	\label{eq:mayer_energy}
	\begin{align}
		&x_{9}(x_{p_{f}}) = \int_{x_{p_0}}^{x_{p_{f}}} \frac{z_T(x_p)}{x_2}dx_p \text{ and}\\
		&x_{10}(x_{p_{f}}) = \int_{x_{p_0}}^{x_{p_{f}}} \frac{z_H(x_p)}{x_2}dx_p. 
	\end{align}
\end{subequations}
Put them together, and we consequently arrive at a cost function with simple Mayer terms
\begin{equation}
\label{eq:opt-time-energy}
J = \alpha x_1(x_{p_{f}})+(1-\alpha)\big(x_9(x_{p_f})+x_{10}(x_{p_f})\big).\\
\end{equation}
This way, the possible issues arising from $x_2$ being zero can be treated consistently throughout the model via the equations defining the dynamics and follows from \cite{barbosa2022time}.

Now the optimization problem involves both the control variables $u(x_p)$ and the auxiliary variables $z_T(x_p)$ and $z_H(x_p)$. Implementation wise though, $u(x_p),~z_T(x_p)$, and $z_H(x_p)$ are treated identically as control inputs. 

However, control inputs and state variables are treated differently in a typical discretization of optimal control problems, causing a degradation of precision. Next we discuss this in more detail and present an approach to alleviate this issue.

\subsection{Auxiliary Parametrization}
\begin{figure}[b]
    \centering
    \vspace*{2.0em}
    \includegraphics[width=0.95\columnwidth]{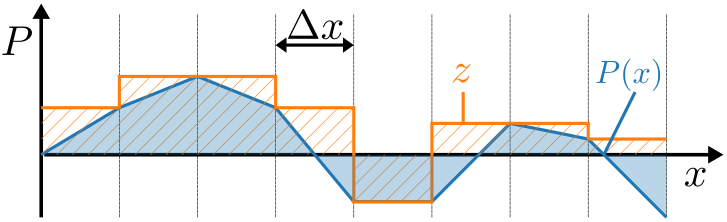}
    \caption{An example of a possible power trajectory where the actual energy consumption is overestimated.}
    \label{fig:powerapproximation}
\end{figure}

With the reparametrization from time to space, the auxiliary parameters $z_T,~z_H$ now need to satisfy
\begin{subequations}
\begin{align}
    z(x_p) &\geq P(x_p) \\
    z(x_p) &\geq \gamma P(x_p),
\end{align}
\end{subequations}
where subscripts $t,h$ are dropped for notational convenience. In a typical discretization of an optimal control problem, states are interpolated as polynomials and control inputs are assumed constant over the control intervals. As such, the function $P(x_p)$ is a polynomial of some finite degree over any given control interval. Additionally, $z(x_p)$ acts as an upper bound for $P(x_p)$, approximating the power over the interval, see Fig.~\ref{fig:powerapproximation}.
Here, a possible power trajectory is visualized in blue and the corresponding auxiliary representation $z$, in orange. The control intervals are visualized as black vertical bars, with width $\Delta x$. Since the piecewise constant $z$ is an upper bound on the power over any given interval, the energy is overestimated and causes a degradation of precision.

To resolve this issue, we represent $z$ as a basis function expansion, i.e., 
\begin{equation}
    z(x_p) = \Phi(x_p)^\top\eta,
\end{equation}
where $\Phi(x_p)^\top=\begin{bmatrix}\phi^1(x_p) & \dots & \phi^K(x_p)\end{bmatrix}$ and $\eta$ are the corresponding weights. In general, $\phi^k(x_p)$ is a problem dependent design choice. In this paper, a direct collocation point based method is used for solving the resulting differential-algebraic system of equations over each control interval. The collocation polynomial degree is chosen as $1$ (i.e., linear) and hence, the state trajectories over each control interval are affine. As such, it suffices to choose a polynomial of degree $1$ for $z(x_p)$.

Hence, in the following, we parametrize $z_H$ over control interval $k$ as
\begin{equation}\label{eq:linearu3}
    z_H^k(x_p) = \eta^k_{H,1} x_p + \eta^k_{H,0},
\end{equation}
where $\eta^k_{H,1}$ and $\eta^k_{H,0}$ are to be optimized over, instead of $z_H$ directly. We use an identical parametrization for the trolley auxiliary variable $z_T(x_p)$. Then, the optimization problem is finally given by
\begin{align}
\label{eq:opt-final}
&\begin{aligned}
& \min_{u, \eta_T, \eta_H}
& & J = \alpha x_1(x_{p_{f}})\!+\!(1\!-\!\alpha)\big(x_9(x_{p_f})\!+\!x_{10}(x_{p_f})\big)\\
& \text{subject to} && \\
\end{aligned}\nonumber\\
&\begin{aligned}
&\, x_{2}x^\prime(x_p) = f(x_p,x(x_p),u(x_p)), && x(0) = x_{0},\\
&\, \eta^j_{T,1}x_p+\eta^j_{T,0}\ge P_T(x_p), && x(x_{p_{f}}) = x_{f},\\
&\, \eta^j_{T,1}x_p+\eta^j_{T,0}\ge \gamma_T P_T(x_p), && 0 \le x_1(x_p),\\
&\, \eta^j_{H,1}x_p+\eta^j_{H,0}\ge P_H(x_p), && 0 \le x_2(x_p),\\
&\, \eta^j_{H,1}x_p+\eta^j_{H,0}\ge \gamma_H P_H(x_p), && u_{min}(x_p) \le u(x_p),\\
&\, g(x_p) \le 0, && u_{max}(x_p) \ge u(x_p),
\end{aligned}
\end{align}
with parameter vectors $\eta_T = \begin{bmatrix} \eta^1_{T,1} & \eta^1_{T,0} & \dots & \eta^K_{T,1} & \eta^K_{T,0} \end{bmatrix}^\top$ and ${\eta_H = \begin{bmatrix} \eta^1_{H,1} & \eta^1_{H,0} & \dots & \eta^K_{H,1} & \eta^K_{H,0} \end{bmatrix}^\top}$.
\section{Simulation Example}\label{sec:results}

To illustrate and validate the idea, a scenario of stack configuration was simulated. For simplicity, a small-scale example was used, where the container stacks are particularly high at the end of the loading site. The optimization problem was modeled using CasADi \citep{Andersson2019} and the Yop toolbox \citep{leek2016optimal} in MATLAB, and IPOPT was the solver used.

All distances and lengths are in \emph{meters} and the angles are in \emph{radians}. With \eqref{eq:opt-final} in mind, the initial and final conditions were set according to
\begin{align*}
    x(0) &= \begin{bmatrix} 0 & 0 & 3 & 0 & 3 & 0 & 0 & 0 \end{bmatrix}^\top\\
    x(x_{p_f}) &= \begin{bmatrix} \hspace{.15em}~ & 0 & 3 & 0 & 3 & 0 & 0 & 0 \end{bmatrix}^\top,
\end{align*}
with $x_{p_0}=0$ and $x_{p_f}=1$. The efficiency of the trolley and the hoisting mechanism were set to $\gamma_T=\gamma_H=0.8$. Finally, the box constraints were set to
\begin{equation*}
\begin{aligned}
    0.15 \le &y_p(x_p) \le h-s(x_p)  \span\span\\
	0 \le &t(x_p)  & 0 \le& \dot{x}_p(x_p)\\
	0 \le &l(x_p) \le 0.75 & -0.1 \le& \theta(x_p) \le 0.1\\
	-1 \le &F_T(x_p) \le 1 & 0 \le& F_H(x_p) \le 8.
\end{aligned}
\end{equation*}

\subsection{Results and Discussion}
\begin{figure}[tb]
\centering
\vspace*{2.0em}
	\includegraphics[width=\columnwidth]{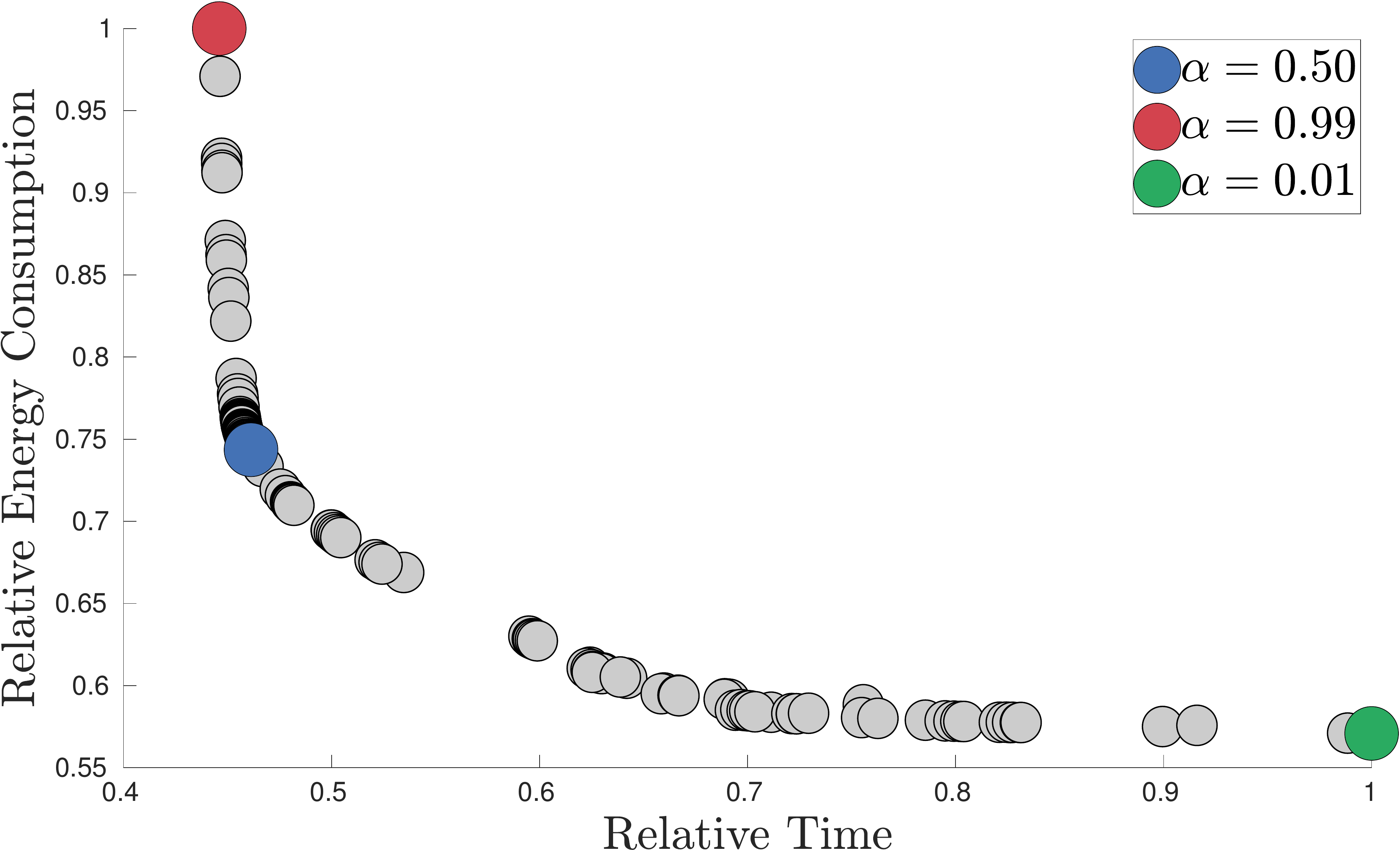}
	\caption{Energy consumption over trajectory time. The energy is relative to the solution with $\alpha=0.99$. The time is relative to the solution with $\alpha=0.01$.}
	\label{fig:paretoplot}
\end{figure}
\begin{figure}[tb]
\centering
	\includegraphics[width=.95\columnwidth]{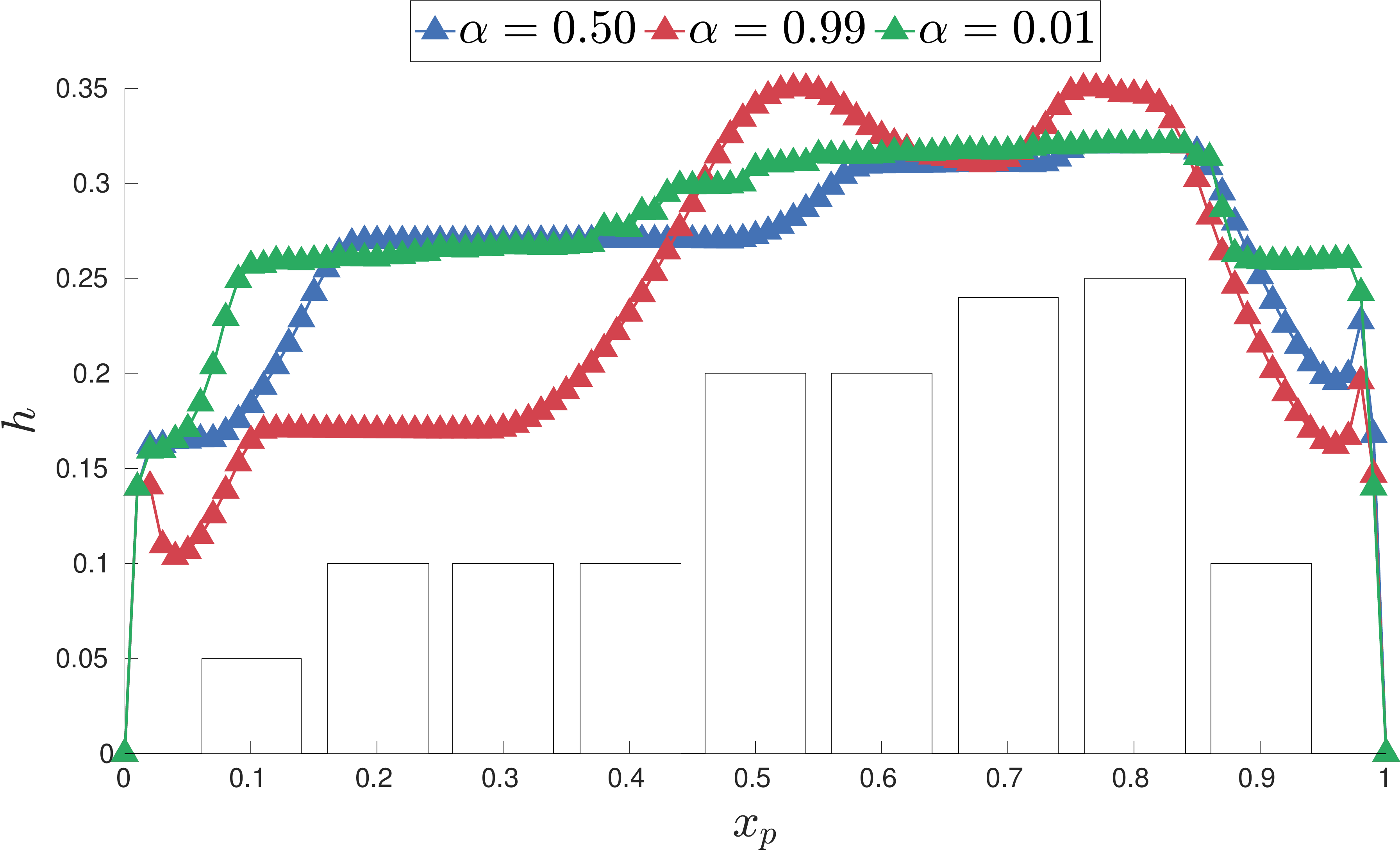}
	\caption{Payload trajectories for the solutions highlighted in Fig.~\ref{fig:paretoplot}.}
	\label{fig:trajplot}
\end{figure}
\begin{figure}[tb]
\centering
	\includegraphics[width=\columnwidth]{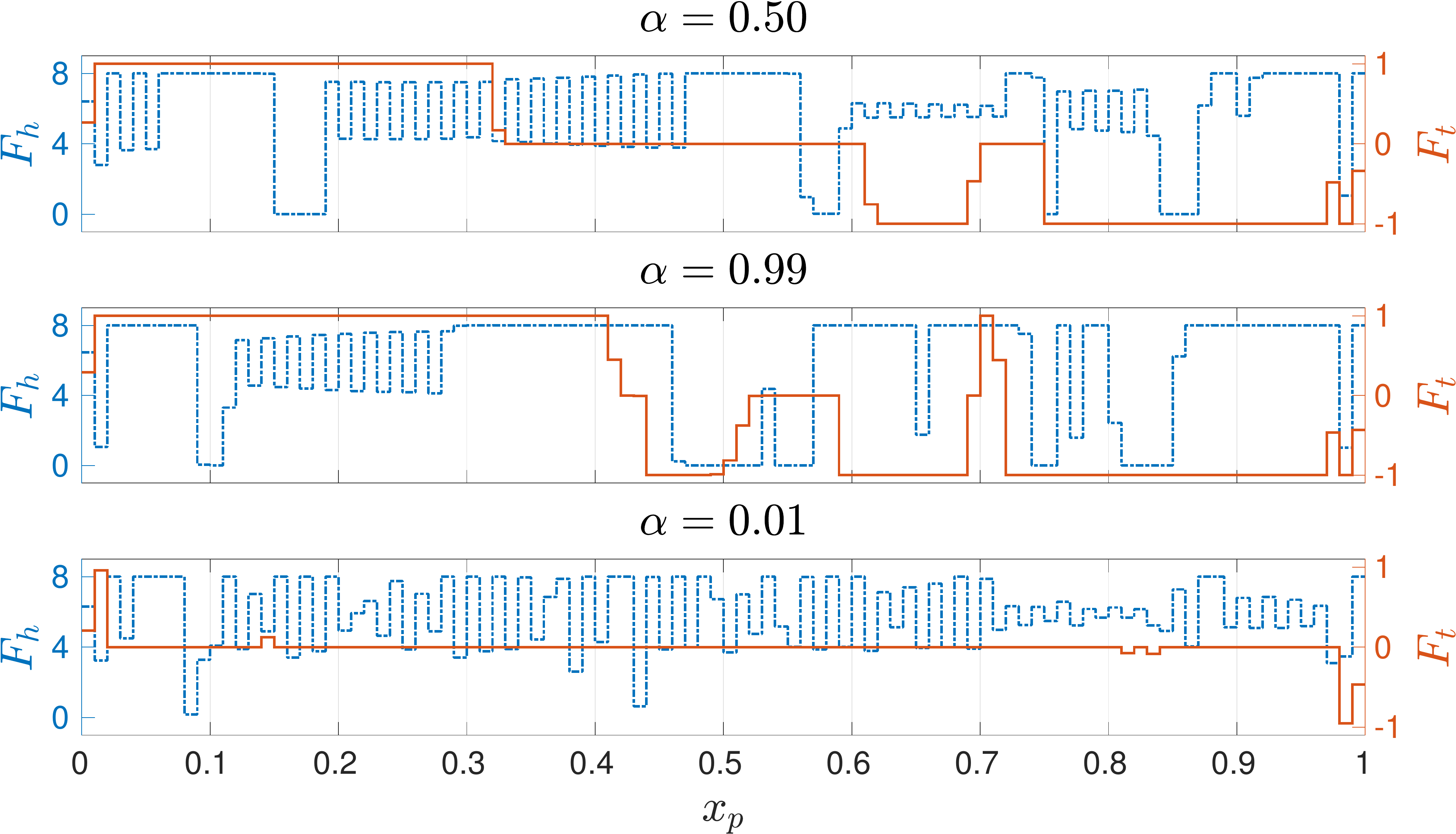}
	\caption{Control inputs for the highlighted solutions.}
	\label{fig:inputplot}
\end{figure}
\begin{figure}[tb]
\centering
\vspace*{2.0em}
	\includegraphics[width=\columnwidth]{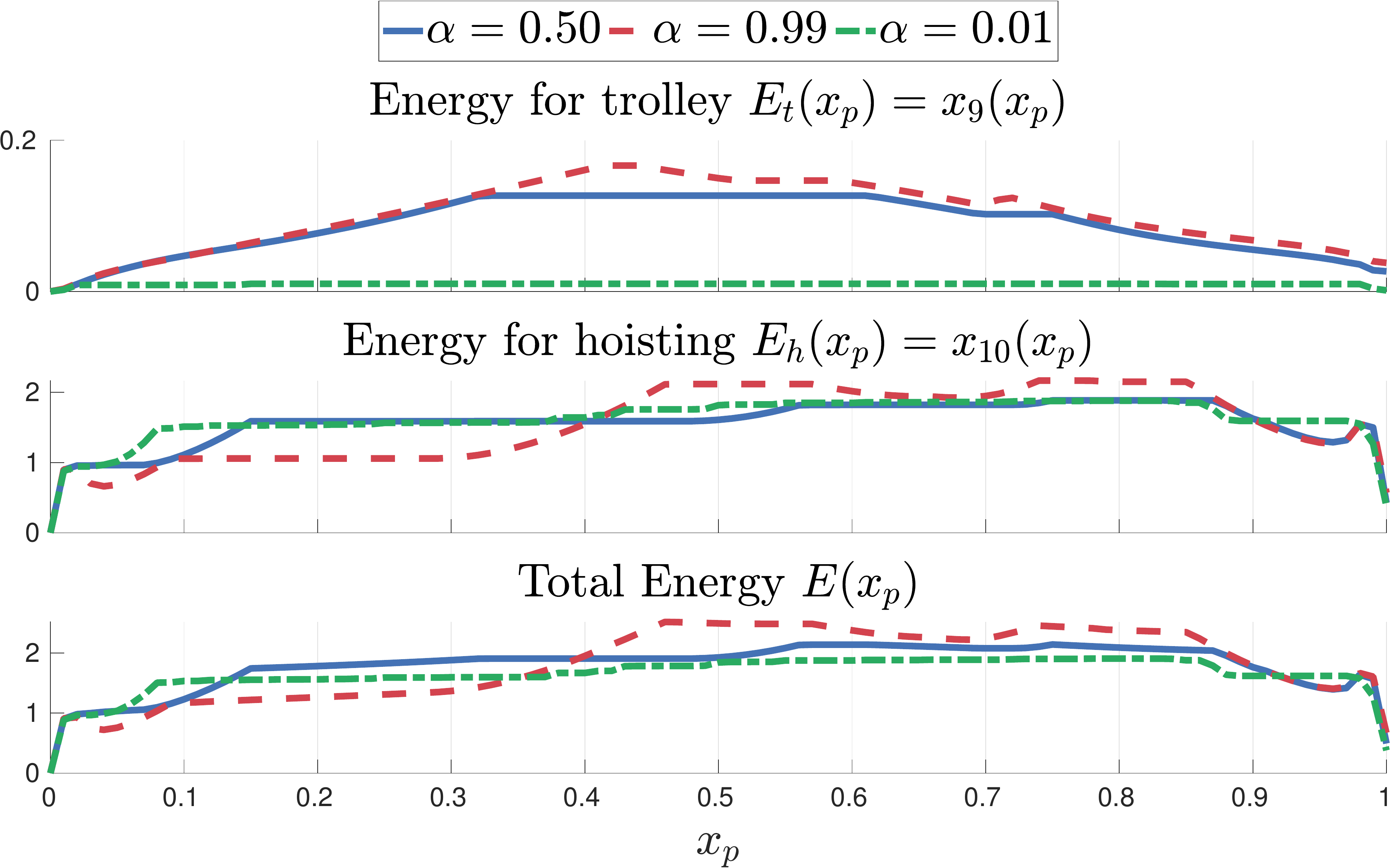}
	\caption{Energy consumption for the highlighted solutions.}
	\label{fig:energyplot}
\end{figure}
\begin{figure}[tb]
\centering
	\includegraphics[width=\columnwidth]{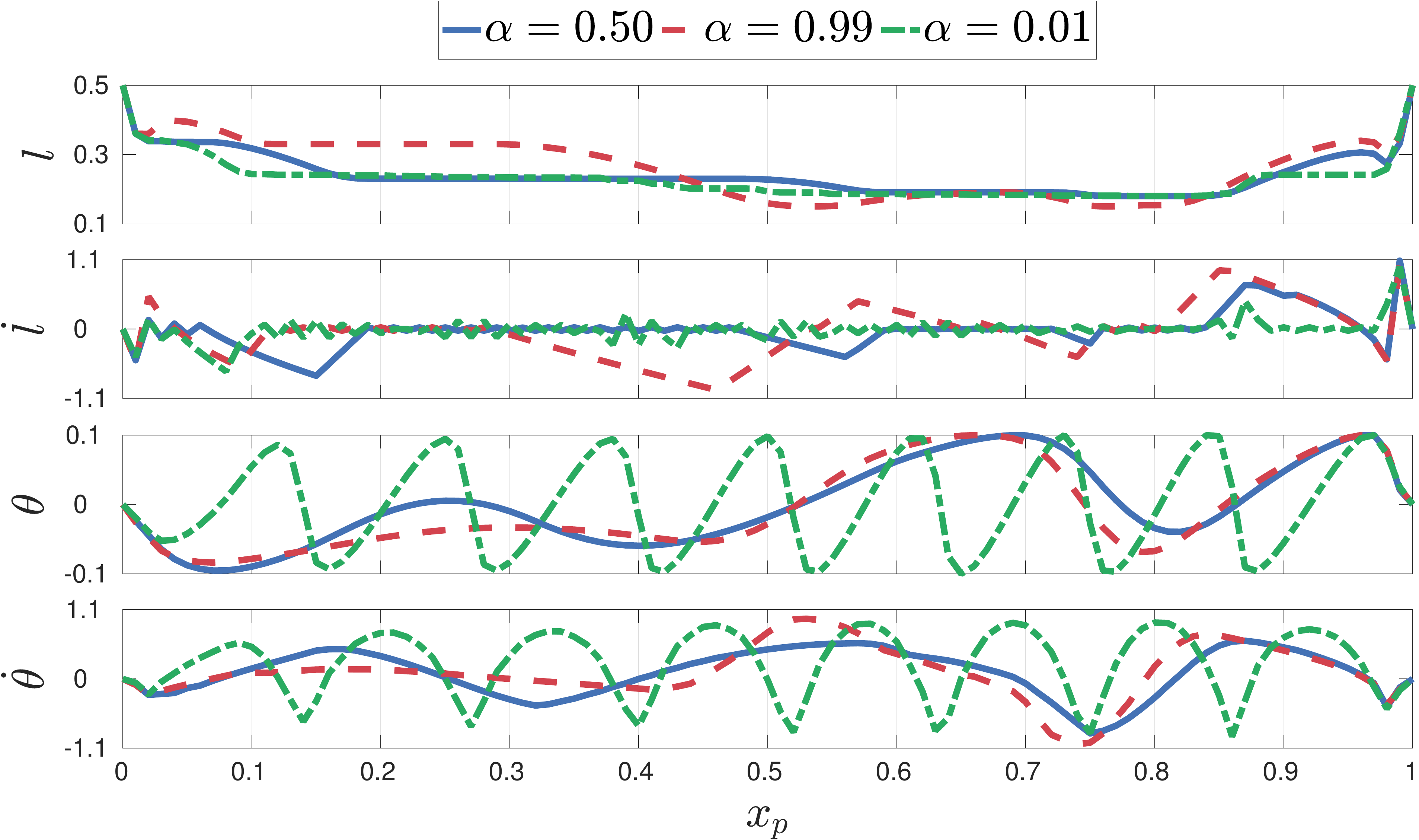}
	\caption{Hoisting and sway for the highlighted solutions.}
	\label{fig:stateplot}
\end{figure}

Fig.~\ref{fig:paretoplot} shows the trade-off between energy consumption and the time for moving the payload to the final state, while avoiding the collision with container the stacks. This was obtained for a range of $0.01 \leq \alpha \leq 0.99$. Three points are highlighted: $\alpha=0.99$, which emphasizes time optimality, $\alpha=0.01$, which emphasizes energy optimality, and $\alpha=0.5$, which has a reasonable trade-off between time and energy consumption. The payload trajectories for the three chosen solutions are visualized in  Fig.~\ref{fig:trajplot}. For these solutions, the applied trolley and hoisting forces are shown in Fig.~\ref{fig:inputplot} and the energy consumption along $x_p$, obtained from $x_{9}(x_{p_f})$ and $x_{10}(x_{p_f})$, in Fig.~\ref{fig:energyplot}. Lastly, the payload hoisting and sway trajectories along $x_p$ are shown in Fig.~\ref{fig:stateplot}.

From Fig.~\ref{fig:paretoplot} one can see that both the approaches minimizing only time or only energy suffer from diminishing returns, and it is evident that there is a minimal amount of time and energy needed to move the payload. The latter is a consequence of the regeneration efficiency not being $\gamma=1$. Moreover, the trade-off solution with $\alpha=0.5$ consumes about $25.6\%$ less energy than the solution with $\alpha=0.99$, while taking only $3.4\%$ longer time. Alternatively, it consumes $30\%$ more energy while taking $54\%$ less time than the solution with $\alpha=0.01$. 

From Fig.~\ref{fig:trajplot} we can see that the solution for ${\alpha=0.99}$ is more aggressive, with more overshoots in the hoisting. This happens because the container momentum helps to propel the trolley forward, which in turn produces higher sway angles $\theta$. The sway is then compensated when the payload is subsequently lowered by the hoisting mechanism, see Fig.~\ref{fig:stateplot}. This is even more evident in the end of the trajectory, when the payload is substantially lowered after avoiding the highest container stack. Energy consumption is affected by this behavior (see Fig.~\ref{fig:energyplot}), with higher peaks and $E(x_{p_f}) = 0.66J$, and more aggressive inputs in Fig.~\ref{fig:inputplot}.

The solution for $\alpha=0.01$ on the other hand consumes a minimal amount of energy to start moving the trolley and recovers it with efficiency $\gamma_T$ when braking. Additionally, the height of the payload is kept constant for longer periods, as small $\vert\dot{l}\vert$ will neither consume energy nor regenerate with efficiency $\gamma_H$. Moreover, note that the payload is early hoisted to a safe height, extending the periods where it is possible to keep nearly constant $l$, see Fig.~\ref{fig:trajplot}. Furthermore, periods of small $\vert\dot{l}\vert$ and oscillatory behavior of $\theta$ in Fig.~\ref{fig:stateplot} suggest transfer of kinetic energy between the payload and the trolley. This is shown in Fig.~\ref{fig:energyplot} where the energy consumption increases rapidly in the beginning and does not vary much during the process until be partially recovered in the end, resulting in $E(x_{p_f}) = 0.38J$. See the control inputs in Fig.~\ref{fig:inputplot}, where $F_T \approx 0$ during most of the process and $F_H$ tries to keep $\vert\dot{l}\vert$ as small as possible. 

Finally, $\alpha=0.5$, the solution chosen as a good trade-off, also shows periods of small $\vert\dot{l}\vert$ as well as trolley acceleration in the beginning, no input in between and regeneration in the end. Note that in this solution, $\theta$ oscillates with lower frequency than for $\alpha=0.01$. Furthermore, the payload trajectory is not as aggressive as for $\alpha=0.99$, with only one overshoot in the very end, yet with  hoisting not as anticipated as for $\alpha=0.01$, see Fig.~\ref{fig:trajplot}. This also results in periods where $E(x_p)$ is nearly constant in Fig.~\ref{fig:energyplot} and $E(x_{p_f}) = 0.49J$.

Note that though alleviated by the reformulations, the optimal control problem remains non-convex, which may lead to local minimum solutions. However, circumventing these limitations is not our goal and non-convexity is a consequence of the high-fidelity model used.

\section{Conclusion} \label{sec:conclusion}

We have studied the trade-off between time and energy consumption for a ship-to-shore crane application with container stack avoidance constraints. For this, we use a simple model to account for energy regeneration, which by simple standard epigraph reformulations can be put in a form with improved numerical properties. Previous work introducing a coordinate system suitable for the application at hand, and allowing the incorporation of container avoidance constraints, was reused when introducing the energy related modeling, allowing for an efficient optimal control problem. By studying a small example, we note that there can be situations where it is possible to substantially reduce energy consumption with a minor increase in loading time. 

Future work will focus on incorporating more complex models of energy regeneration.

\bibliography{ifacconf}             
\end{document}